\documentclass[lettersize,journal]{IEEEtran}
\usepackage{amsmath,amsfonts}
\usepackage{algorithmic}
\usepackage{algorithm}
\usepackage{array}
\usepackage[caption=false,font=normalsize,labelfont=sf,textfont=sf]{subfig}
\usepackage{textcomp}
\usepackage{stfloats}
\usepackage{url}
\usepackage{verbatim}
\usepackage{graphicx}
\usepackage{cite}
\usepackage{diagbox}
\begin{document}

\title{HCB: Enabling Compact Block in Ethereum Network with Secondary Pool and Transaction Prediction}

\author{\IEEEauthorblockN{ Chonghe Zhao,Taotao Wang, \IEEEmembership{Member, IEEE}, Shengli Zhang, \IEEEmembership{Senior Member, IEEE},  Soung Chang Liew, \IEEEmembership{Fellow, IEEE}}
	\IEEEcompsocitemizethanks{\IEEEcompsocthanksitem Chonghe Zhao is with the College of Electronics and Information Engineering, Shenzhen University, Shenzhen 518060, China  \protect\\(e-mail: zhaochonghe\_szu@163.com).
		\IEEEcompsocthanksitem Taotao Wang and Shengli Zhang  are with the College of Electronics and Information Engineering, Shenzhen University, Shenzhen 518060, China, and also with MAXE LAB (e-mail: zsl@szu.edu.cn; ttwang@szu.edu.cn).
		\IEEEcompsocthanksitem Soung Chang Liew is with the Department of Information Engineering, The Chinese University of Hong Kong, Hong Kong SAR, China, and also with MAXE LAB (e-mail: soung@ie.cuhk.edu.hk).
	}
	\thanks{Corresponding author: Shengli Zhang}
}



\maketitle

\begin{abstract}
Compact block, which replaces transactions in the block with their hashes, is an effective means to speed up block propagation in the Bitcoin network. The compact block mechanism in Bitcoin counts on the fact that many nodes may already have the transactions (or most of the transactions) in the block, therefore sending the complete block containing the full transactions is unnecessary. This fact, however, does not hold in the Ethereum network. Adopting compact block directly in Ethereum may degrade the block propagation speed significantly because the probability of a node not having a transaction in the sending block is relatively high in Ethereum and requesting the missing transactions after receiving the compact block takes much additional time. To investigate the factors that prevent compact block in Ethereum, we set up probe nodes to collect data from Ethereum MainNet and performed data analysis. Our analysis results indicate that the missing transactions could be attributed to factors such as small transaction pools, network latency, and miners' selfish behaviors. Moreover, simply enlarging the transaction pool and using the prediction algorithm proposed for Bitcoin to predict the missing transactions and prefetch them do not work for Ethereum. This paper proposes hybrid-compact block (HCB), an efficient compact block propagation scheme for Ethereum and other similar blockchains. First, we develop a Secondary Pool to store the low-fee transactions, which are removed from the primary transaction pool, to conserve storage space. As simple auxiliary storage, the Secondary Pool does not affect the normal block processing of the primary pool in Ethereum. Second, we design a machine learning-based transaction prediction module to precisely predict the missing transactions caused by network latency and selfish behaviors. We implemented our HCB scheme and other compact-block-like schemes (as benchmarks) and deployed a number of worldwide nodes over the Ethereum MainNet to experimentally investigate them. Experimental results show that HCB performs best among the existing compact-block-like schemes and can reduce propagation time by more than half with respect to the current block propagation scheme in Ethereum. 
\end{abstract}


\section{Introduction}\label{intro}
\IEEEPARstart{E}{thereum}, regarded as the architype of Blockchain 2.0, introduced Ethereum Virtual Machine (EVM) and smart contract to fulfill various Turing-complete computing tasks in a decentralized manner \cite{ethereum}. Taking a step beyond Bitcoin, which only implements a distributed ledger to record transactions \cite{bitcoin}, Ethereum can support various distributed applications (DAPPs) in the domains of  Metaverse \cite{metaverse}, Decentralized Finance (DeFi) \cite{mev}, and Non-fungible Token (NFT) \cite{NFT}. Ethereum is becoming the most popular blockchain system with around 10,000 active nodes and a cryptocurrency market capitalization approaching that of Bitcoin.

With more and more DAPPs deployed on Ethereum, the low number of transactions per second (TPS) Ethereum can process is becoming a significant performance bottleneck. According to \cite{etherscan}, the current TPS of Ethereum is 15, and around 200,000 pending transactions are usually piled up in the network. 

Quite a number of works tried to improve the TPS performance of public blockchains (i.e., Bitcoin-like and Ethereum-like blockchain) from the perspectives of consensus, sharding, and networking \cite{performanceBlockchain}. The aim of this work is to improve the TPS of Ethereum-like public blockchains from the networking perspective.   A straightforward but naïve way to scale TPS is to embed more transactions into each block. However, a block of large size is encumbered with large block propagation time and increased fork rate \cite{fork1}. Compact block, which compresses the block size by obviating the need to propagate the transactions already in the receiving node's transaction pool (Tx-Pool), is an effective way to increase the number of transactions encapsulated into each block. In a compact block, the transactions in the full block are replaced with their hashes, which the receiving node uses to identify the transactions from within its local transaction pool \cite{compact}. Some transactions may be missing in the local Tx-Pool, in which case extra communication rounds are needed for the receiving node to obtain the missing transactions. Since the block size of Bitcoin is large (up to 1MB), compact block has been widely studied and adopted \cite{graphene,xthinblock,blockEntropy,txilm}. Graphene \cite{graphene} uses a Bloom Filter \cite{bloom} and Invertible Bloom Lookup Tables (TBLT) \cite{IBLt} to compress blocks further and to quickly detect missing transactions. XThin \cite{xthinblock} and XThinner \cite{blockEntropy} combine Bloom Filter and shorter transaction hash to decrease missing transactions and compress block. Txilm \cite{txilm} optimizes the size of the transaction hash by evaluating the probability of hash collisions and introduces a “salt” in the hash computation to protect against potential attacks.

Compact block potentially could improve the TPS of Ethereum. There are two main differences between the Ethereum and Bitcoin blockchain regarding the block propagation process in the networking layer. First, Ethereum adopts a hybrid block-hash propagation protocol that simultaneously forwards the full blocks and block hashes over the network. The sending node forwards the full block to a random subset of its neighbor nodes and the block hash to its other neighbor \cite{ethna}. Seldom does a receiving node that receives the block hash already has all the transactions in the block unless it has already received the full block from another sending node. Thus, nodes mostly obtain new blocks through the reception of full blocks. This means that full-block broadcasting dominates the block propagation in Ethereum, the same as in Bitcoin prior to the introduction of compact block. Second, the block size of Ethereum is smaller than that of Bitcoin. With the growth of Ethereum gas limit, its block size is increasing, and most blocks in Ethereum are larger than 20KB, as shown in \cite{etherscan}. As demonstrated in \cite{information,scale}, the block propagation time increases quickly when the block size exceeds 20KB. Moreover, our work in \cite{bbp} also shows that a small-sized “bodyless block” can improve block propagation speed by up to 4 times in Ethereum.  

Designing an efficient and compatible compact block scheme for Ethereum is challenging due to some unique mechanisms of Ethereum. According to \cite{bbp} and \cite{churn}, the matched-block probability is around 0.85 in Bitcoin, but only around 0.12 in Ethereum. In other words, for Ethereum, about 88\% of the compact blocks contain missing transactions at a receiving node, and the need for extra rounds of communication to obtain the missing transaction reduces the efficiency of compact block significantly. Improving the matched-block probability is paramount to the effectiveness of compact block in Ethereum. Predicting missing transactions and including the predicted missing transactions into the compact block is a good way to improve matched-block probability. Works \cite{prediction ,confirmation} investigate the transaction prediction problem for Bitcoin, and work \cite{compact} applies transaction prediction to the compact block scheme in Bitcoin. In \cite{compact}, when forwarding a compact block, the sending node first predicts that the receiving nodes are likely not to have some of the transactions conveyed by this compact block, the sending node then piggybacks these missing transactions with the compact blocks and sends them together with the compact block to the receiving nodes. Meanwhile, \cite{prediction} uses a Random Forest Classifier \cite{randomForest} to decide which transactions in the Mempool (transaction pool in Ethereum) to pack into the next block; \cite{confirmation} compares multiple classifiers to predict the confirmation time of the transactions in the Bitcoin network. Compared with Bitcoin, the transaction pool and block interval in Ethereum are much smaller, and a new compact block scheme must be compatible with these system setups (simply changing the system parameters may induce other issues that are not compatible with the whole system). Furthermore, new DAPPs bring new behaviors to Ethereum when propagating transactions, which also need to be taken into account. For example, some transactions related to DeFi smart contracts might be withheld within a few nodes for privacy reason or for higher profit \cite{flashbots} (i.e., preferences are given to other transactions with higher fees). 

In this paper, we propose a hybrid-compact block (HCB) protocol to speed up block propagation in Ethereum. Unlike the entirely new compact block scheme in \cite{bbp} that is hard to work in the current Ethereum without modifying the functional components of block validation and assembly, HCB is fully compatible with the current Ethereum. In HCB, we introduce a Secondary Pool module and a Missing Transactions Prediction module to increase the matched-block probability of compact blocks in Ethereum.

Our main contributions are summarized as follows:
\begin{itemize}
	\item Identifying the causing factors of the subpar matched-block probability in Ethereum: We set up probe nodes to collect data from Ethereum MainNet. Our investigation on the collected data shows that the poor matched-block probability in Ethereum is caused by small transaction pools, network latency, and nodes' selfish behaviors. We quantify the relative contributions of the three factors to the poor matched-block probability. We prove that with the current Ethereum architecture, merely enlarging the transaction pool would result in a larger empty block rate, reducing the TPS performance.

	\item Proposing a secondary transaction pool (Secondary Pool): Instead of simply enlarging the transaction pool, we propose a Secondary Pool module to store transactions removed from the original small transaction pool because of their relatively low fees. The transactions in the Secondary Pool can also contribute to the reconstruction of the full block when an HCB block is received. We design a transaction restore algorithm that returns selected transactions in the Secondary Pool to the original transaction pool after each block.  
	
	\item Predicting the missing transactions: We apply machine-learning model of Naïve Bayes Classifier to build a Missing Transactions Prediction model and train the model by collecting data from Ethereum MainNet. To decrease the retransmission probability, HCB piggybacks the missing transactions predicted by the model and other transaction hashes into the compact block for forwarding. The precision of our prediction model can reach 0.951.
	
	\item Implementing HCB and Testing over the Ethereum Mainnet: We implement an HCB prototype and conduct experiments over the Ethereum MainNet with a set of nodes located worldwide to evaluate the performance of our HCB prototype. We compare the experimental performance of HCB with the current Ethereum's block propagation protocol and three other compact block-like propagation protocols. The experimental results show that the matched-block probability of HCB in Ethereum is around 0.90, which is even better than that of compact block in Bitcoin. Notably, the block propagation time of HCB is only about half that of the current Ethereum. 
\end{itemize}

The rest of this paper is organized as follows. Section \ref{experimentEther} measures and presents the performance of basic compact blocks in Ethereum. An overview of our hybrid-compact block protocol is given in Section \ref{overview}. Section \ref{specificdesign} elaborates on the specific design of our hybrid-compact block protocol. Section \ref{experimentpre} designs the experiments and discusses the experimental results. Section \ref{conclusion} concludes this work.

\section{Basic Compact Block Protocol in Ethereum}\label{experimentEther}
We set up an experiment to test the basic compact block protocol in Ethereum. We found that applying the basic compact block protocol in Ethereum leads to even worse performance than the current full block protocol due to the high retransmission rate induced by the low matched-block probability. Identifying the causes and the remedy for the low matched-block probability leads us to the design of our HCB protocol.

\subsection{Poor Performance of Basic Compact Block Protocol in Ethereum}\label{experimentsta}

Although the basic compact block protocol can speed up the block propagation in Bitcoin \cite{compactblockbenefits}, it may not work well in Ethereum. As shown in our experiment result in \cite{bbp}, the matched-block probability in Ethereum is only around 0.12 compared with 0.85 in Bitcoin \cite{churn}. In other words, if Ethereum simply incorporates the basic compact block protocol into its block propagation process, the nodes would often need extra rounds of communications to acquire the missing transactions.

\begin{figure}[tp]
	\centering
	\includegraphics[width=8cm]{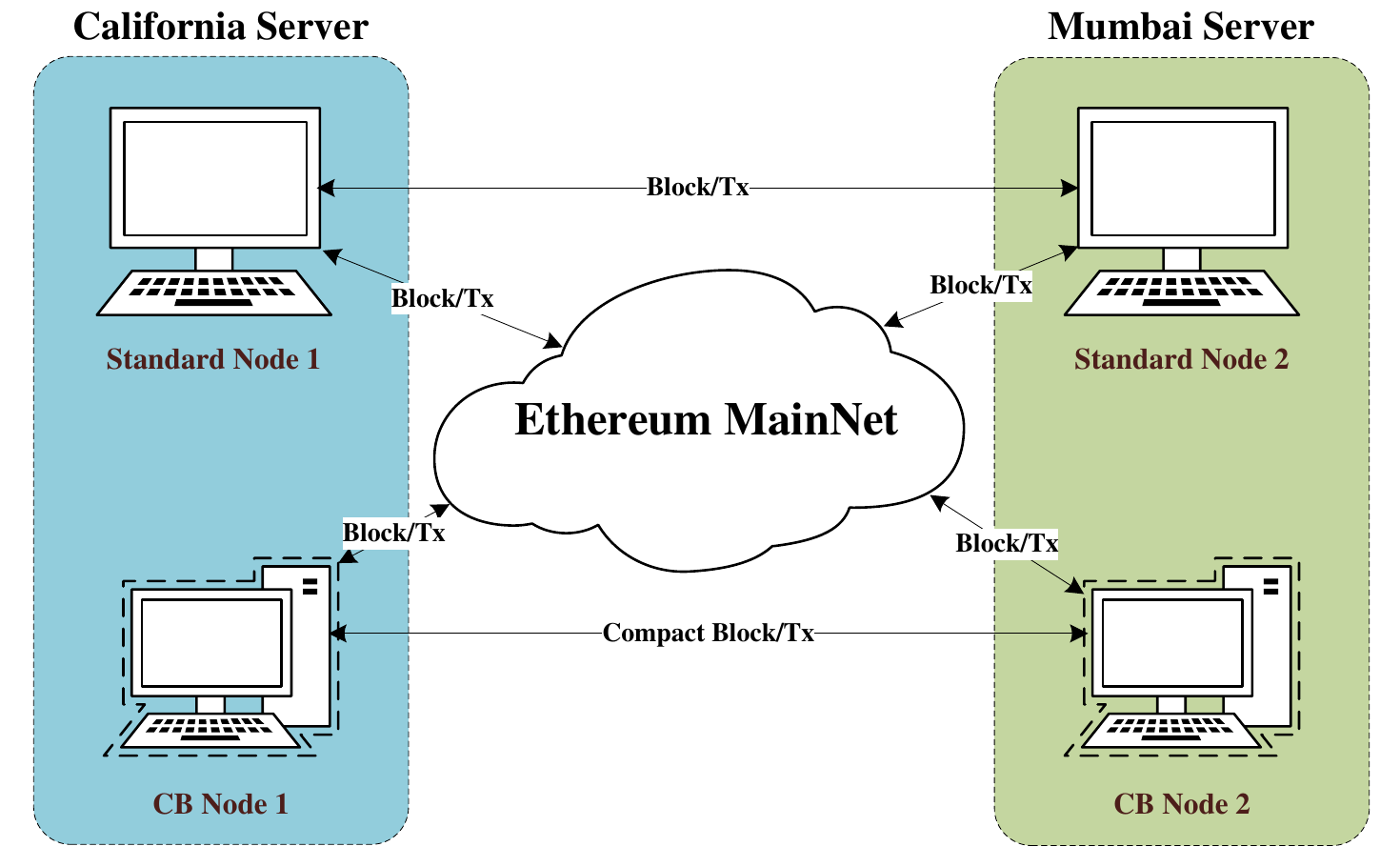}
	\caption{The experimental setup for investigating the performance of the basic compact block protocol in Ethereum. }
	\label{compactExperiment}
\end{figure}

To gain insight into compact block propagation in Ethereum, we set up an experiment in Ethereum MainNet. We compare the transmission time of full blocks and that of compact blocks plus the transmission times of missing transactions. As shown in Fig. \ref{compactExperiment}, we set up two standard full nodes (Standard Node 1 and Standard Node 2) that propagate full blocks like other legal Ethereum nodes would do and two modified full nodes (CB Node 1 and CB Node 2) that propagate compact blocks instead of full blocks on two AliCloud servers\footnote{One of the AliCloud servers was located in California with an IP address of 47.251.1.118, and another one was located in Mumbai with an IP address of 149.129.181.111. Both the AliCloud servers have the same typical hardware configuration: CPU with 8 cores, RAM with 16GB, SSD with 1.5 TB, and bandwidth with 8 Mbit/sec.}. Each server contains one standard node and one CB node: one server contains Standard Node 1 and CB Node 1, and the other server contains Node 2 and CB Node 2. The two standard full nodes run with the Ethereum Geth client \cite{GoEthereum}, and the two modified full nodes run with the modified Ethereum Geth client that incorporates the basic compact block protocol. All the four nodes were connected to Ethereum MainNet, and the two standard nodes and the two CB nodes were connected respectively, during the period of August 2, 2022 to August 6, 2022. 

The specific experimental process is as follow: Standard Node 1 and Standard Node 2 exchange the transactions and full blocks with Ethereum MainNet. Standard Node 1 and Standard Node 2 forward the full blocks to each other. That is, the Standard Node that receives a full block would first forward it to the other Standard Node directly. Similarly, CB Node 1 and CB Node 2 also exchange the transactions and blocks with Ethereum MainNet. To forward compact blocks, the CB Node that first receives a full block from Ethereum MainNet would transform the received full block to a compact block and then sends it to the other CB node. The receiving CB Node would reconstruct the full block with its local transactions. If the receiving CB Node cannot match all CB transactions in its local transaction pool, it requests transmissions of the missing transactions. 

\begin{figure}[tp]
	\centering
	\includegraphics[width=7.5cm]{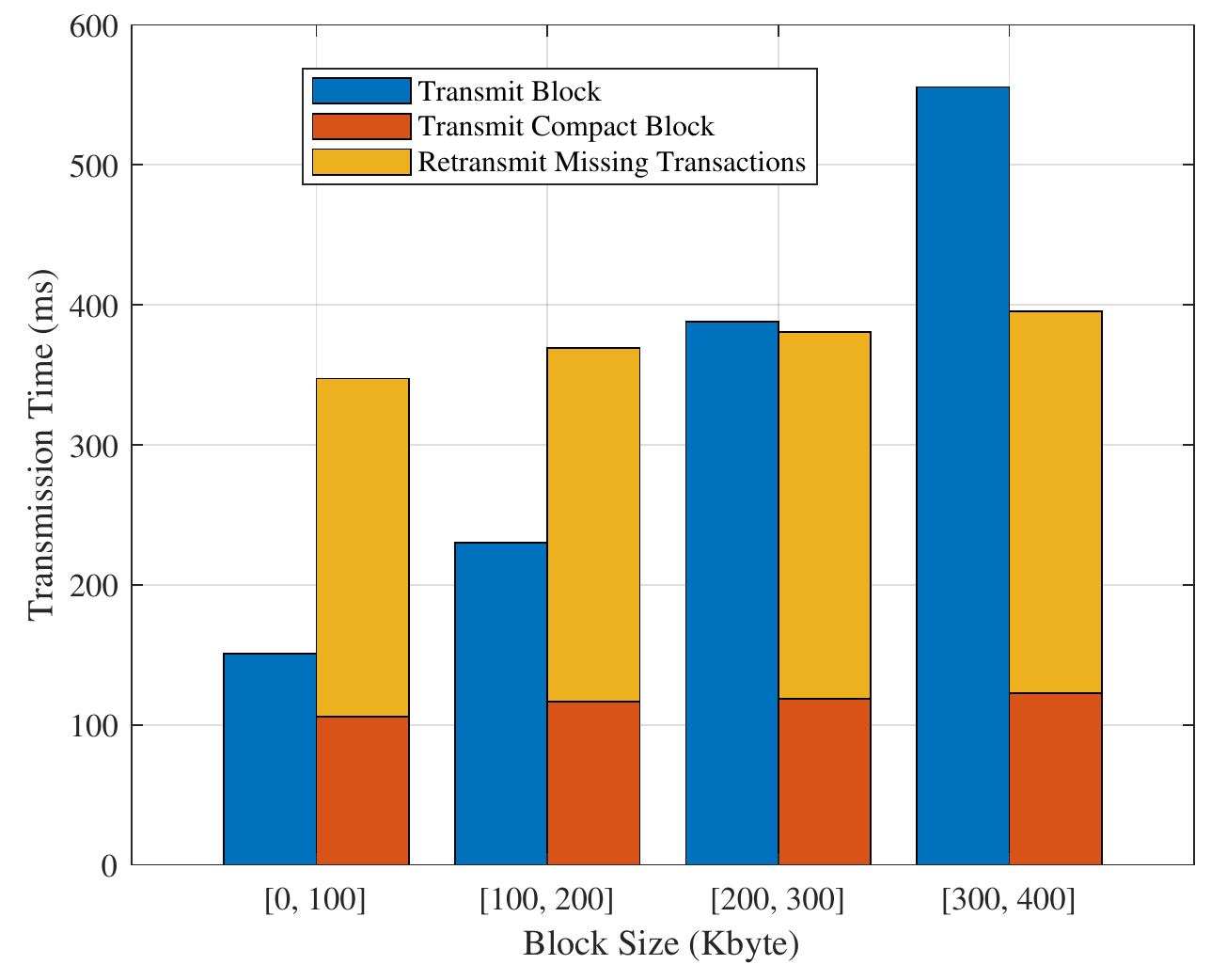}
	\caption{The measured average transmission time of full blocks (blue), compact blocks (red), and the measured average retransmission time of missing transactions (yellow) in our experiment. }
	\label{compactReTransmission}
\end{figure}

We analyze the data recorded in the log files of the four nodes. We first measured the average transmission time of the recorded 28987 full blocks transmitted between Standard Node 1 and Standard Node 2. We then measured the average transmission time of the corresponding recorded 28987 compact blocks transmitted between CB Node 1 and CB Node 2. Among the 28987 compact blocks, 26279 compact blocks contain some missing transactions at the receiving CB Nodes. The matched-block probability in our experiment is 0.11, close to the measured results in \cite{bbp}. Fig. \ref{compactReTransmission} plots the average block transmission time of the two protocols for different full block sizes. The blue bars are the average transmission time of the full blocks, which increases linearly with the block size. The average transmission time of compact blocks consists of the initial compact block transmission time   (red bars) and the missing transactions transmission time (yellow bars). The initial compact block transmission time nearly keep constant for different full block sizes, since different full block sizes do not change the corresponding compact block sizes much; the missing transactions transmission time increase very slowly with the block size  due to the slow increase in the number of missing transactions with the block size. When the block size is smaller than 200Kbytes, the performance of the basic compact block protocol is worse than the full block protocol. When the block size is larger than 200Kbytes, even though the CB nodes also need missing transactions to be transmitted with high probability, the performance of the compact block protocol is better than that of the full block protocol, whose transmission time is dominated by the block size. Additionally, according to \cite{etherscan} and our experiment, most of the blocks are smaller than 200Kbytes in the current Ethereum MainNet, and thus the basic compact block protocol does not work well. To improve the performance of the basic compact block protocol, we need to increase the matched-block probability (hence reducing the retransmissions of missing transactions).

\subsection{Factors Causing Low Matched-block Probability}\label{factorcas}

To identify the causes of the low matched-block probability of the basic compact block protocol in Ethereum, we analyze and classify the missing transactions observed in our experiment. We recorded all the transactions in the received compact blocks that were not at the receiving CB nodes. Two sets $S_1$ and $S_2$ record the information about the missing transactions at CB Node 1 and CB Node 2, respectively. Each entry $\{hash,t_r,t_c\}$ in $S_i$ corresponds to one missing transaction in CB Node $i$, where hash is the hash of the missing transaction, $t_r$ is the time of reception of the transaction from the Ethereum MainNet (i.e., this transaction has been removed from the node's Tx-Pool due to some reason before the arrival of the compact block that contains the hash of the transaction), and $t_c$ is the time of reception of the transaction hash in the compact block. Notably, if the CB node never received a missing transaction, $t_r$ is $NULL$. By analyzing the missing transactions in $S_1$ and $S_2$, we identify three types of missing transactions caused by different factors. 

\begin{itemize}
	
	\item \textbf{Stale transactions caused by small transaction pools (67.45\%):} A stale transaction is defined as a missing transaction whose $t_r !=NULL \&\&  t_r<t_c$. In other words, a stale transaction was once received and stored in the node's local transaction pool, but it had been removed from the transaction pool before the compact block arrived. The stale transactions were removed due to the limited transaction pool size to make room for the numerous transactions arriving later. According to \cite{etherscan}, there are around 200,000 pending transactions piled up over Ethereum MainNet, and a CB node that runs with the default transaction pool of the Geth client can only store 6144 transactions in its transaction pool. Thus, CB nodes remove old transactions to make room for the new transactions according to the transaction pool evicting algorithm in Ethereum \cite{GoEthereum}. The inclusion of a stale transaction in a block occurs when an old transaction is included in a compact block by the miner that operates with an extended-size transaction pool. Our experiment shows that the proportion of stale transactions among the missing transactions at CB Node 1 and CB Node 2 are 66.7\% and 68.2\%, respectively. 
	
	\item \textbf{Selfish transactions caused by miners' selfish behaviors (32.20\%):} 	A selfish transaction is defined as a missing transaction whose $t_r==NULL$, which means that selfish transactions were never received by the receiving nodes when the compact block arrived. In the Ethereum MainNet, some full nodes deliberately withhold some transactions without forwarding them to other nodes and pack these transactions directly into their mined blocks. For example, the users who participate in the FlashBots Project only send their transactions related to DeFi contracts to some specific miners without broadcasting these transactions to the whole network of miners \cite{flashbots}. Our experiment shows that the proportion of selfish transactions among the missing transactions at CB Node 1 and CB Node 2 is 32.9\% and 31.5\%, respectively. 
	
	\item \textbf{Later transactions caused by network latency (0.35\%):} A later transaction is defined as a missing transaction whose $t_r !=NULL \&\& t_r>t_c$, which means that a later transaction arrives at the receiving CB node after the compact block that contains this transaction is received by this receiving CB node. This later arrival of transactions is caused by the network latency when propagating transactions over Ethereum MainNet \cite{ethna}. Typically, the receiving time of later transactions were within 1 second after the corresponding compact blocks were received $(t_r-t_c<1 s)$, and on average were around 548ms in our experiment. The experiment result shows that the later transactions are 0.4\% and 0.3\% of the total missing transactions at the two CB nodes, respectively. 
	
\end{itemize}
\section{HCB Overview}\label{overview}
As discussed in the last section, we must deal with the three types of missing transactions to improve the matched-block probability so that the compact block protocol can be adopted in Ethereum. Intuitively, we can enlarge the size of transaction pool to solve the problem of stale transactions, and we can use transaction prediction and piggyback to solve the problems of selfish and later transactions. However, these intuitive ideas cannot be directly applied in Ethereum because they may have the opposite effect of performance degradation, as will be explained shortly. This section first discusses the technical challenges for applying compact block protocol in Ethereum and then gives the overview our HCB solutions.

\subsection{Technical challenges}
\textbf{Challenge 1: } Simply enlarging the size of transaction pool increases the empty block rate, and thus decreases the TPS performance in Ethereum. 

In Ethereum, after receiving a new block from other nodes, a miner node will immediately begin to mine an empty block that contains no transactions while verifying the received block to update its local global state \cite{GoEthereum}. The reason to mine an empty block is that the miner needs to first reset its local transaction pool before assembling the next new block, during which time the miner can only mines an empty block. If the hash puzzle for the empty block is solved before a block with transactions can be assembled, then the miner successfully mines the empty block and the miner will broadcast the empty block over the network. On the other hand, if the hash puzzle for the empty block has not been solved by the time a block with transactions is assembled, then the miner will switch to mining the new block. 

We now quantitatively model the relationship between TPS and the size of transaction pool in Ethereum. The miner executes two processes after receiving a new block form another node: the local transaction pool reset and the new block assembly, both of which depend on the size of the miner's local transaction pool. When executing the transaction pool reset, the miner first removes the invalid transactions after the blockchain global state is updated by the new block (e.g., with the transactions packed in the newly received block, some transactions with insufficient balances after the blockchain global state is updated). After that, the miner needs to re-calculate the Gas fees of the transactions in the pool and re-rank the transactions according to the re-calculated Gas fees, since Gas fees in EIP-1559 are dynamically determined based on the updated blockchain global state \cite{EIP1559}. When assembling a new block, the miner packs and executes the eligible transactions from the local transaction pool according to the priority of transaction selection and ordering in Ethereum \cite{bbp}. 

\begin{figure}[tp]
	\centering
	\includegraphics[width=7cm]{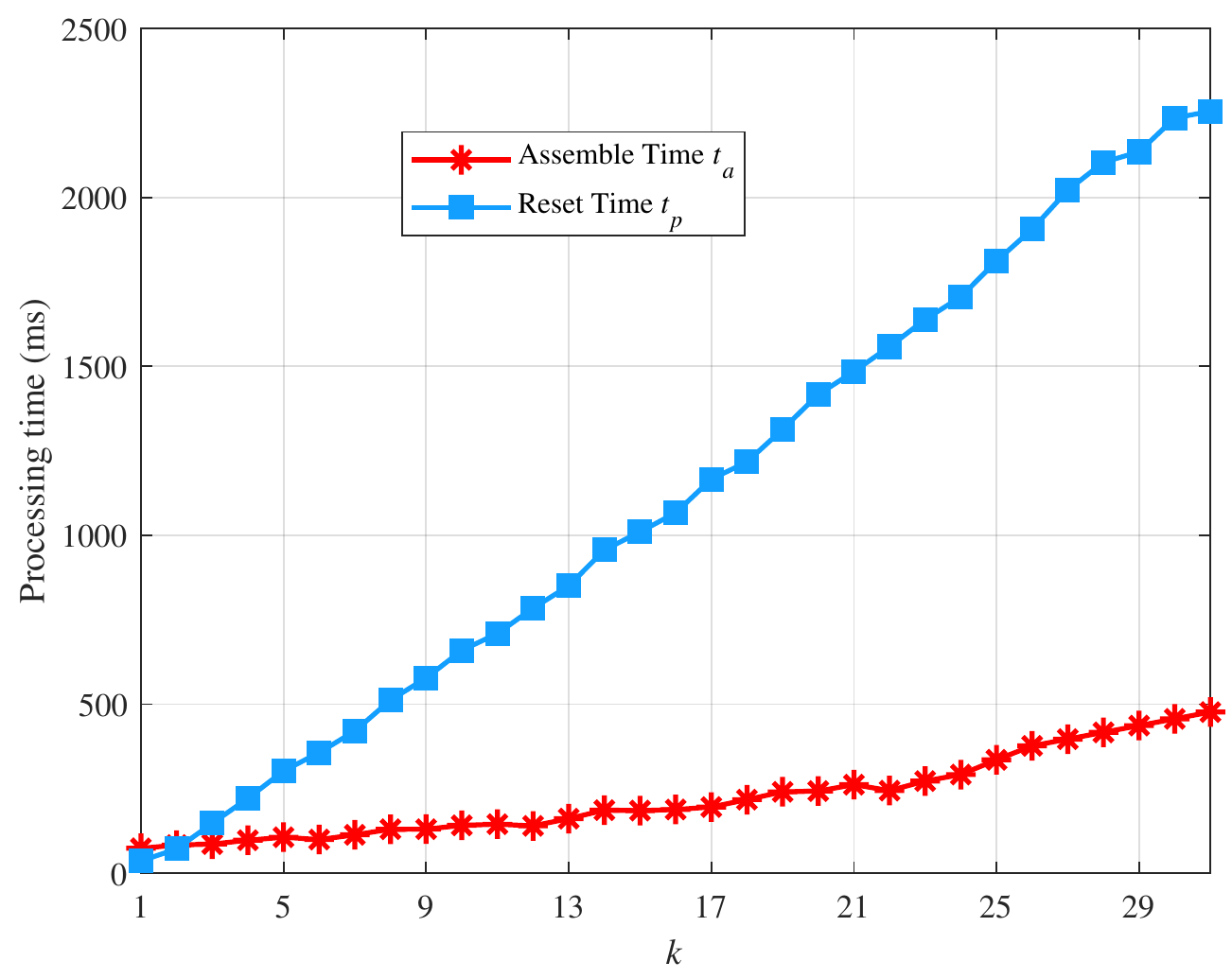}
	\caption{The measured results of the block assembly duration $t_a$ and the transaction pool reset duration $t_p$ under different $k$. }
	\label{minerAssembly}
\end{figure}

In the PoW blockchain systems, such as Bitcoin and Ethereum 1.0, the block generation rates follow the exponential distributions \cite{transactionFee,bloxroute,blockFee}. Given that the average block interval in Ethereum is measured as 13000ms in \cite{etherscan}, we model the Ethereum's block interval to follow an exponential distribution with a mean of $t_g=$ 13000ms. Further, we denote the duration of transaction pool reset by $t_p$, and the duration of the new block assembly by $t_a$. Then, the duration during which a miner is mining an empty block is $\tau=t_p+t_a$, and the probability that an empty block is successfully mined in this duration is given by:
\begin{equation}
	p(\tau ) = 1 - {e^{\frac{{ - \tau }}{{{t_g}}}}} = 1 - {e^{\frac{{ - ({t_p} + {t_a})}}{{{t_g}}}}}
	\label{forktransaction}
\end{equation}
Considering that there are on average $M$ transactions in a full block, the effective TPS in Ethereum can be computed as 
\begin{equation}
	TPS = (1 - p(\tau ))*\frac{M}{{{t_g}}} = {e^{\frac{{ - ({t_p} + {t_a})}}{{{t_g}}}}}*\frac{M}{{{t_g}}}
	\label{TPS}
\end{equation}
Note that, in eq. (\ref{forktransaction}), both $t_p$ and $t_a$ increases with the size of the transaction pool. We then set up an experiment to measure $t_r$ and $t_a$ under different sizes of the transaction pool during the period of March 12, 2022 to March 15, 2022. We denote the default size of the transaction pool by $L$, and we have $L=6144$ (i.e., there are up to 6144 transactions in the transaction pool by default). In our experiment, we gradually enlarge the size of the transaction pool to $k*L,k=1,2,\cdot  \cdot  \cdot,$   and measure the values of $t_p$, $t_a$ when the transaction pool is full of transactions. The measured results are plotted in Fig. \ref{minerAssembly}. From the measurement results, we can observe the linear increases of $t_p$, $t_a$  with the size of the transaction pool (in terms of k in Fig. \ref{minerAssembly}). Therefore, we can fit two linear functions to model the corresponding relationships: $t_a=12.9k+56.5$ and $t_p=73.9k-37.9$. Thereby, using these linear functions, we now can rewrite eq. (\ref{forktransaction}) as 
\begin{equation}
	p(k) = 1 - {e^{ - \frac{{86.8k + 18.6}}{{{t_g}}}}}
	\label{forkcurrent}
\end{equation}
The accuracy of (\ref{forkcurrent}) can be verified by setting $k=1$ as in Ethereum: the empty block rate given by our model (\ref{forkcurrent}) is $p(k=1) \approx 0.01$; and it is shown in \cite{etherscan} that the real empty block rate in Ethereum is also around $0.01$. With (\ref{forkcurrent}), if miners enlarge the transaction pool to store up to 200,000 transactions ($k=33$), the empty block rate is $p(33) \approx 0.20$; and thus, the TPS will decreases from 15.3 to 12.3 (there are around average $M=200$ transactions in a full block). Therefore, it is not a good way to solve the stale transactions by simply enlarging the size of transaction pool. 

\textbf{Challenge 2:} It is difficult to precisely predict selfish and later transactions in Ethereum, since the network latency is dynamic and there are many complex applications, causing the snapshots of the transaction pools of different nodes to be quite different. 

To improve the matched-block probability of the compact block protocol, a sending CB node can predict the missing transactions at receiving CB nodes and piggybacks the predicted missing transactions along with the sent compact block. One way to predict selfish transactions is to use Gas fees of transactions, i.e., predicting the transactions with high Gas fees as the selfish transactions. One way to predict later transactions is to use ages of transactions (the age of a transaction is defined as the time duration since the CB node received the transaction), i.e., predicting the transactions with small ages as the later transactions. However, it is difficult to determine the optimal age to predict the later transactions, since the latency when propagating transactions over the network is dynamic and it is also related to Gas fees (some nodes may discard the transactions with low Gas fees and withhold the transactions with high Gas fees, which influence the propagation latency). Also, the complex applications deployed over Ethereum may render the prediction of selfish transactions ineffective. Notably, Ethereum supports DeFi applications by smart contracts, where miners' actual profits not only come from Gas fees, but also from Miner Extractable Value (MEV) by including, excluding or reordering some specific transactions in blocks \cite{mev}, e.g., a transaction with low Gas fee may call the internal instruction of a smart contract to pay extra MEV for miners privately \cite{flashbots}. Therefore, it is infeasible to simply predict the selfish transactions based on Gas fees and the later transactions based on ages. 

\begin{figure}[tp]
	\centering
	\includegraphics[width=7cm]{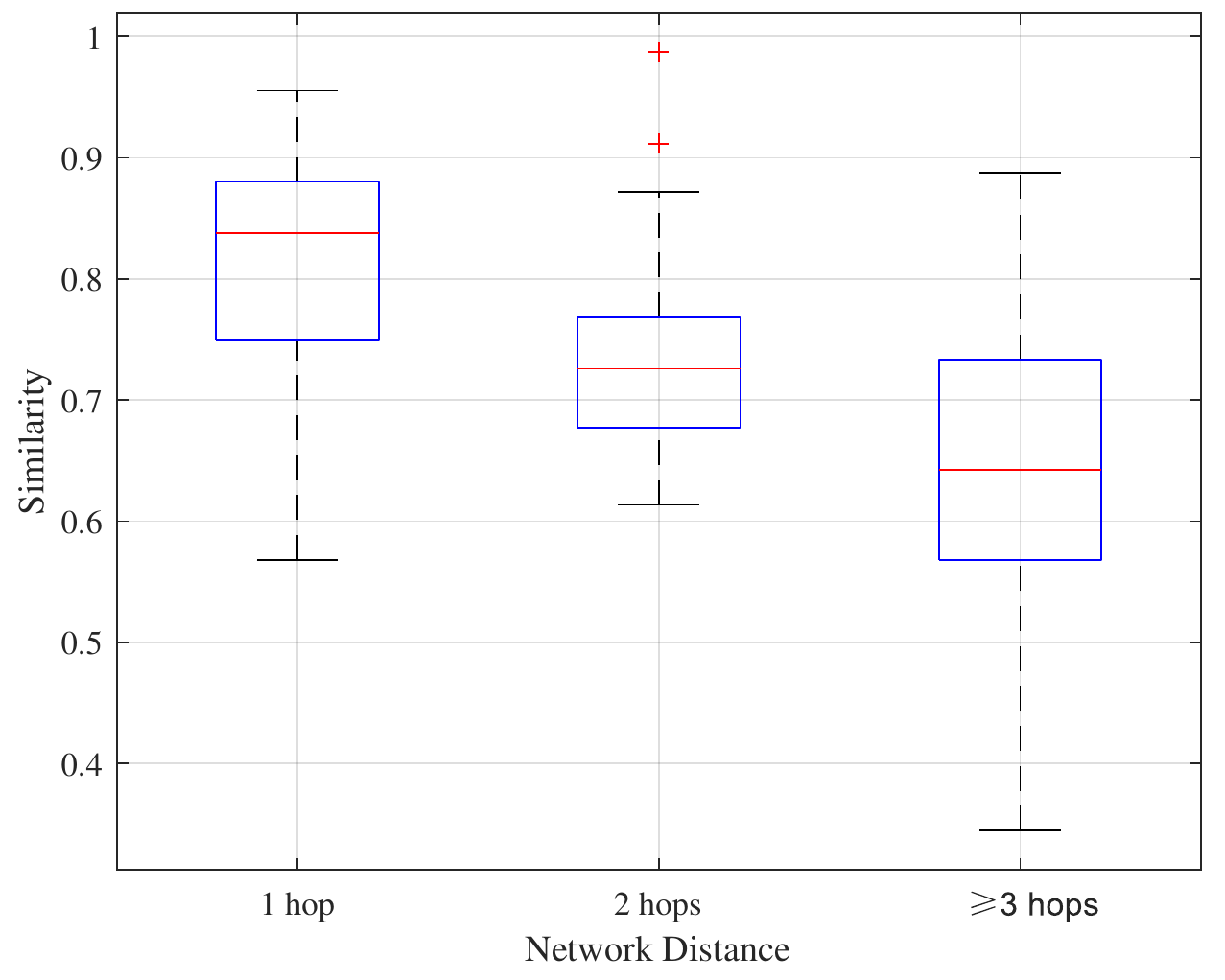}
	\caption{The similarity of the transaction pools between the two experimental nodes with respect to different network distance.  }
	\label{similaritytxpool}
\end{figure}

Another way to predict missing transactions is based on the local transaction pool: when a CB node receives a block, it first checks to see which transactions are in the block but not in its local transaction pool, and then predicts these transactions as the missing transactions for other nodes and piggyback them with the compact block. The performance of this prediction method depends on the similarity of transaction pools, which is good in Bitcoin \cite{compactFAQ}. If the similarity of transaction pools between different nodes is good, it is very likely that a transaction not in the transaction pool of a sending node is not in the transaction pool of a receiving node either. However, this simple prediction method does not work well if the transaction pools of different nodes are not similar. Therefore, we measured the similarity of transaction pools in Ethereum MainNet. We set up two standard full nodes that connect to Ethereum MainNet in the three days from March 9, 2022 to March 11, 2022. The two standard full nodes are both set up with the same configuration as the Standard Node 1 and Standard Node 2 in the experiment described in the last section. On the first day, we directly connected the two nodes and thus the network distance between them was one hop.  On the second day, we did not directly connect the two nodes but connected them to a common neighbor node, and thus their network distance was two hops. On the last day, we removed the direct connection between the two nodes and also removed their connections to the common neighbor node, and thus their network distance was at least three hops. We define the similarity of two transaction pools as the number of the common transactions in the two pools over the number of the transactions in the union of the two pools. We measured the similarity of the two transaction pools of the two experimental nodes by comparing their transaction pools every 10 minutes. Fig. \ref{similaritytxpool} shows the measured similarity of the two transaction pools with respect to different network distance. Although the median similarity can reach 0.84 for the case of one hop network distance, the network distances between the majority of Ethereum nodes are not one hop. Moreover, the median similarity is only around 0.64 when the network distance is more than two hops. For comparison, \cite{txpoolsimilarity} shows that the similarity of transaction pools in Bitcoin is around 0.997 in most cases. Based on our experimental results, we can conclude that the prediction method based on the transaction pool would not work well in Ethereum.
\subsection{Overview of HCB Solution}
\begin{figure}[tp]
	\centering
	\includegraphics[width=8.5cm]{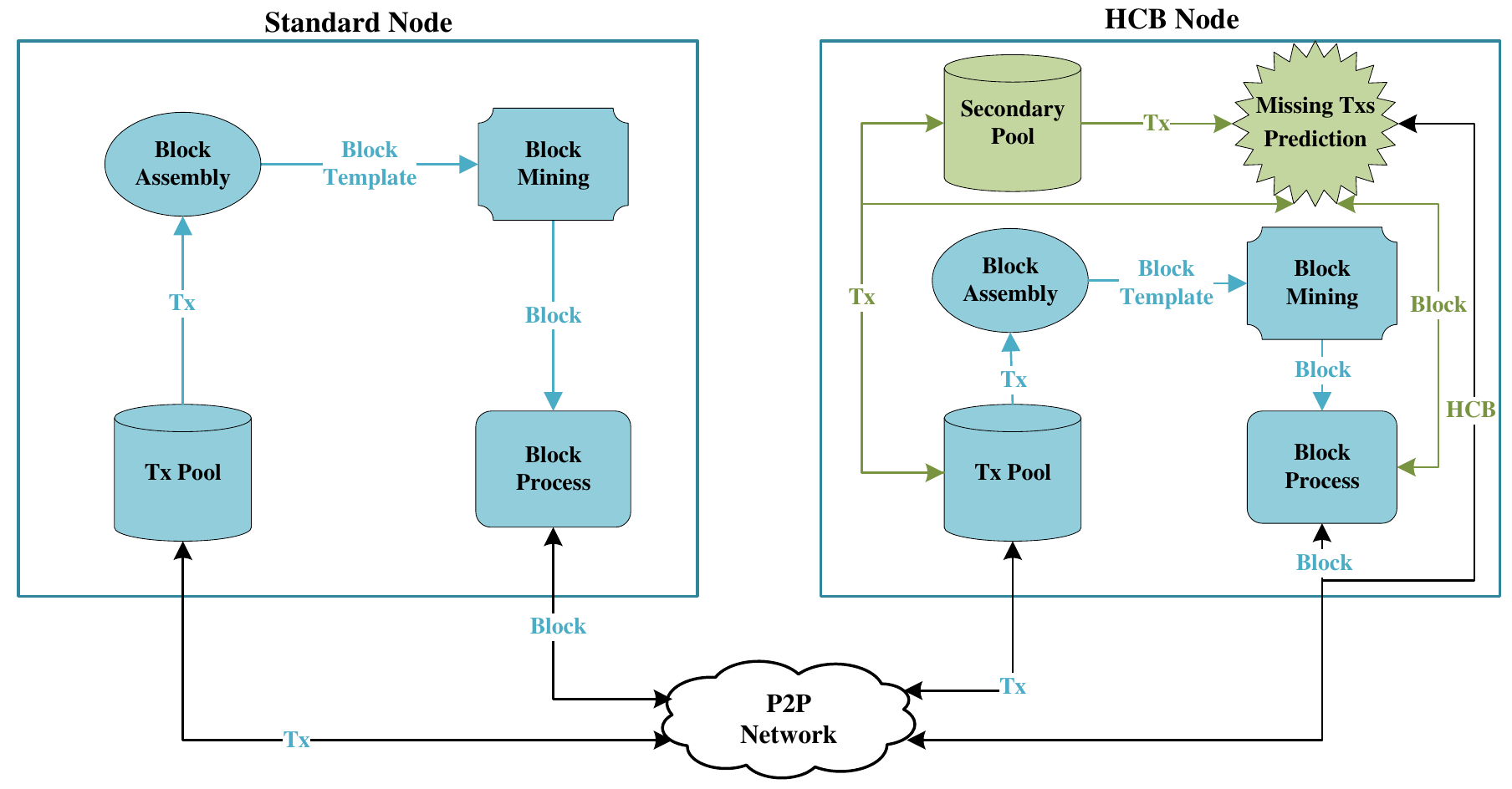}
	\caption{The architectures of standard Ethereum node and our HCB node.   }
	\label{hcbarchitectures}
\end{figure}

To tackle the above challenges, we propose a hybrid-compact block (HCB) protocol to deal with the three types of missing transactions and speed up block propagation for Ethereum. Our HCB is fully compatible with the protocol stack of the current Ethereum blockchain. To implement HCB, we extend the present Ethereum implementation with the architecture shown in Fig. \ref{hcbarchitectures}. There are four original modules in a standard Ethereum node: Transaction Pool (Tx-Pool), Block Assembly, Block Mining, and Block Process. A standard Ethereum node receives new blocks and transactions from other standard nodes over the underlying P2P network; the Tx-Pool module stores new and unpacked transactions; the Block Assembly module selects eligible transactions from Tx-Pool to assemble new blocks; the Block Mining module computes mining puzzles to generate legal blocks with valid nonces; the Block Process module validates new blocks (mined locally or received from other nodes), updates the local database, and forwards the legal blocks to other nodes.   

Extending the standard Ethereum node, our proposed HCB nodes add a Secondary Pool module and a machined learning based Missing Transactions (TXs) Prediction module to enable the compact block protocol. The aim of Secondary Pool is to solve the problem of stale transactions: specifically, Secondary Pool stores the transactions that are removed from Tx-Pool due to their low Gas fees \footnote{Gas fee of a transaction is dynamic over time after EIP-1559 is adopted by Ethereum. Therefore, it will be removed from Tx-Pool to Secondary Pool when its Gas fee is low. If its Gas fee increases significantly after a certain time, it can also be recovered to Tx-Pool from Secondary Pool like a new transaction. }. When an HCB block is received, we can reconstruct the full block by using the transactions from both the Tx-Pool and the Secondary Pool. In our HCB protocol, the standard Tx-Pool is not modified (its size and function); the block mining module still select eligible transactions from Tx-Pool to assemble a new full block. Therefore, the empty block rate is not increased either, resolving Challenge 1. The aim of Missing TXs Prediction module is to solve the problems of selfish and later transactions by predicting and piggybacking these transactions into HCB blocks, so as to decrease the transaction retransmission probability. We train a data-driven prediction model of naïve Bayes classifier with the empirical transaction data sampled from Ethereum MainNet. Although the similarity of Tx-Pools in Ethereum is small, we improve the prediction precision by considering various transaction features related to the network latency and miners' selfish behaviors \footnote{Using our data-driving prediction framework, more complex and advanced prediction scheme can be applied to improve the precision further.}. 

\begin{figure}[tp]
	\centering
	\includegraphics[width=8.5cm]{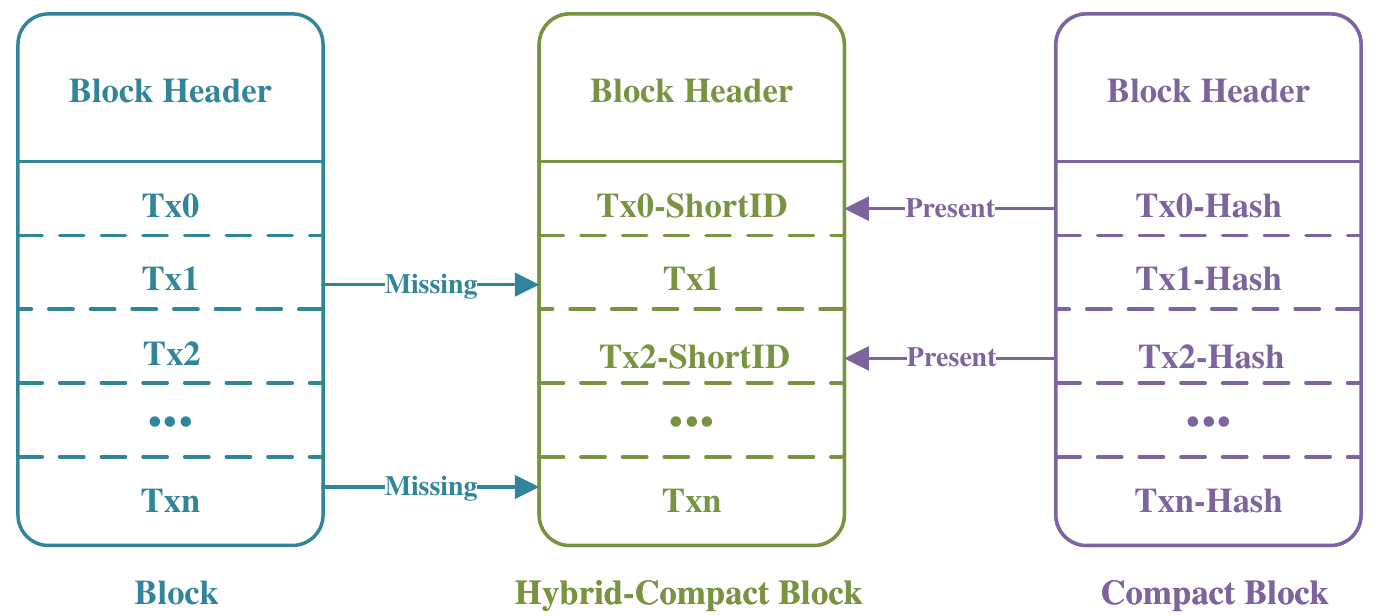}
	\caption{The structure of a standard block, an HCB block, and a compact block.  }
	\label{hcbcompact}
\end{figure}

An HCB block can be regarded as a combination of a standard block and a compact block. The data structure of HCB blocks is shown in Fig. \ref{hcbcompact}. Each transaction in an HCB block is classified as missing or present with the help of the Missing TXs Prediction module. For a missing transaction, the full transaction data is included into the HCB block. For a present transaction, its short hash is packed into the HCB block to reduce overload (a 6 bytes short ID is used in our scheme \cite{compact}).  

Our HCB protocol can work with the current Ethereum with full compatibility. Based on the extended architecture in Fig. \ref{hcbarchitectures}, an HCB node is able to generate HCB blocks from full blocks, and vice versa. An HCB node uses Missing TXs Prediction to generate an HCB block. It then forwards the HCB block to other HCB nodes and the full block to other standard nodes that do not support our HCB protocol.  The receiving HCB node can use the full transactions, which are predicted to be missing at the receiving HCB node and included in the HCB block by the sending node, and the transactions in its local Tx-Pool and Secondary Pool to reconstruct the full block.

\section{Detailed HCB Design} \label{specificdesign}
This section presents the specific design of HCB, including the new modules of Secondary Pool and Missing Transactions Prediction, and the HCB forwarding protocol. 

\subsection{Secondary Pool}
In HCB, each node maintains a Secondary Pool to solve the stale transactions, and the Secondary Pool interacts closely with the original Tx-Pool (primary).

Before introducing the Secondary Pool, we first briefly review the mechanism of the original  Tx-Pool for Ethereum nodes. There are two submodules in Tx-Pool: Pending and Queue. Pending stores the transactions with continuous nonces (i.e., the transactions issued from the same account must contain the continuous nonces), which will be selected to assemble the full block \cite{ethna}. The default size of Pending is 5120. Queue stores the transactions with discontinuous nonces, and its default size is 1024. To avoid the broadcast of invalid transactions over the network, the transactions must be verified successfully before being stored in Tx-Pool, including signature, balance, size, and so on \cite{wood2014ethereum}. After verification, the valid transactions are stored in Pending or Queue according to the values of their nonces, and only the transactions stored in Pending are forwarded. At the same time, all transactions in Tx-Pool are ordered by their fees. When the number of transactions in Tx-Pool reaches the full value, i.e., 5120 plus 1024, the new transactions with higher fees evict the old ones with lower fees (i.e., these evicted transactions are stale transactions discussed in Section \ref{factorcas}),     and also the positions of the transactions in Pending and Queue will be readjusted according to their nonces. 

As a complement to the primary Tx-Pool, our Secondary Pool has a dynamic size and three main processes for maintenance, i.e., the processes of transaction storage, transaction identification and block-wise update, as explained in the following: 

\textbf{Dynamic Size: }Since the secondary pool size does not affect the empty block rate, it can be set to a large value to store large stale transactions to improve the matched-block probability. The only concern is memory consumption, especially if the attackers issue a transaction spam attack to deliberately send a lot of valid but low fees transactions to HCB nodes. Thus, we place a dynamic limit on the secondary pool size, which is synchronized once a day with the number of unconfirmed transactions over the Ethereum network published on the website of \cite{etherscan}. For example, it is set to 200,000 on July 6, 2022. The reason to set this dynamic limit on the secondary pool size is that we assume that the average number of unconfirmed transactions will not change suddenly and we will not accept any more transactions beyond the used the secondary pool size in case of a spam attack. 

\textbf{Transaction Storage: }A stale transaction to be removed from Tx-Pool due to the arrival of other transactions would be stored in the Secondary Pool if space permits. If the secondary pool is full, the transaction with lowest fee would be discarded. A stale transaction received for the first time with continuous nonce but cannot be stored in Pending due to its low fees (i.e., it is evicted by some coming transaction with higher fee from Pending, and thus it is a stale transaction now) would be stored in the Secondary Pool and would also be forwarded to other nodes. 

For fast reconstruction of the full block and the handling of the hash collision (multiple transactions with the same short ID), we set up a short ID map to record the transactions in Tx-Pool and the Secondary Pool. The short ID map consists of pairs of key-value, where each key is a short ID of transactions and the corresponding value is an array containing the transactions whose short IDs equal to that key. A transaction and its short ID are stored into the short ID map as new pair of key-value when its short ID has not been recorded in the map; otherwise, it is appended to the value of an existing key-value pair when its short ID has been recorded in the map. Compared with the size of the standard hash, the size of short ID is reduced from 32 Bytes to 6 Bytes in our scheme. According to \cite{compact}, this short ID will cause a slightly increased probability of hash collisions; however, it is acceptable for our HCB protocol\footnote{In our experiments, there was no any hash collision occurring within one week, as shown in Section \ref{experimentpre}.}. Notably, to support our HCB protocol, the short ID map also needs to record the transactions in Tx-Pool and their short IDs to reconstruct the full block.

\textbf{Transaction Identification: }When a node receives an HCB block, it identifies the present and missing transactions by querying the short ID map about the short IDs of the transactions contained in the HCB block. If the short IDs of the transactions are contained in the short ID map, these transactions are present ones; otherwise, they are missing transactions in our HCB protocol. Then the node uses the present transactions to construct the full block if there are no missing transactions. If there are transactions missing from both of Tx-Pool and the secondary pool, the node sends the corresponding short IDs of the missing transactions to the neighbor node that forwarded  this HCB block to request the missing transactions. At the same time, the node also records the identification result of each transaction (i.e., present or missing) as one of the features for subsequently Missing Transactions Prediction. 

Specifically, the node adopts the following rules to identify the transactions in the HCB block. For each short ID in the HCB block, if it cannot be found in the keys of the short ID map, the corresponding transaction is identified as missing; moreover, if it can be found in the short ID map as a key and its value (the array) only contains one transaction, the transaction is identified as the only transaction corresponding to this short ID (i.e., having no hash collision). There is a little probability that  the value (the array) corresponding to a given key contains multiple transactions (i.e., having hash collision). We propose the following algorithm to resolve the hash collision problem in our HCB.  

To solve the hash collision problem that causes the unnecessary transaction request and retransmission, we add a post-identification phase processing in our HCB protocol. After the identification process of all transactions in the HCB block, we use a set $U_m$ to store the corresponding short IDs of the transactions that are identified as missing, a set $U_p$ to store the corresponding short IDs of the transactions that are uniquely identified precisely (the no hash collision case), and a set $U_c$ to store the short IDs that each correspond to multiple transactions (the hash collision case). If $U_m$ is not empty, the node identifies the transactions in sets $U_c$ and $U_m$ as missing, and requests these missing transactions from the sending nodes of this HCB block.  If $U_m$ is empty and $U_c$ is non-empty, the node can try all possible combinations of the collided transactions to check which one satisfies the value of the $BodyHash$\footnote{$BodyHash$ is a standard hash computed from all transactions in the block body.}  field in the block header. 

\textbf{Block-wise Update: }After reconstructing the full block, the node validates the full block and updates its local database to obtain the new global states. Next, the Secondary Pool will be reset based on the new global states. During the reset operation, first, all invalid transactions are removed from the Secondary Pool, including the transactions with insufficient balance and the expired transactions that are already contained in the block. Then, according to the base fee and gas fee used in the block header, the fees of the remaining transactions are updated by the design principle of EIP-1559 \cite{EIP1559}. Finally, we propose a transaction restore algorithm to move the transactions whose fees increase significantly under the new global states, to Pending and Queue. The proposed transaction restore algorithm is given in Algorithm \ref{restore}.

\begin{algorithm}[t]
	\caption{Transaction Restore Algorithm} 
	\label{restore}

	\begin{algorithmic}[1]
		
		\REQUIRE  a sequence $G_p$ storing the transactions in Pending of Tx-Pool by ascending order of fees; a sequence $G_q$ storing the transactions in Queue of Tx-Pool by ascending order of fees; a sequence $G_s$ storing the transactions in the Secondary Pool by descending order of fees.                  
		\ENSURE $G_p$, $G_q$, $G_s$
		\STATE Select the first transaction $Tx_p$ with the lowest fee ${\gamma _p}$ in $G_p$.
		\STATE Select the first transaction $Tx_q$ with the lowest fee ${\gamma _q}$ in $G_q$.
		\STATE Select the first transaction $Tx_s$ with the highest fee $\beta $ in $G_s$.
		\STATE $\gamma  = \min ({\gamma _p},{\gamma _q})$
		\WHILE{$\gamma<\beta$} 
		\IF{(the nonce of $Tx_s$ is continuous for $G_p$) \&\& ($\gamma_p<\beta$)} 
		\IF{the limit of Pending is reached} 
		\STATE Remove $Tx_p$ from $G_p$ and insert it in $G_s$.
		\STATE Remove the transactions with discontinuous nonces from $G_p$ and insert them in $G_s$.
		\ENDIF
		\STATE Remove $Tx_s$ from $G_s$ and insert it in $G_p$.
		\STATE Update $Tx_p$, $Tx_s$, $\gamma_p$, $\gamma$, and $\beta$.
		\ELSIF{(the nonce of $Tx_s$ is discontinuous for $G_p$) \&\& ($\gamma_q<\beta$) }
		\IF{the limit of Queue is reached}
		\STATE Remove $Tx_q$ from $G_q$ and insert it in $G_s$.   
		\ENDIF
		\STATE Remove $Tx_s$ from $G_s$ and insert it in $G_q$.
		\STATE Update $Tx_q$, $Tx_s$, $\gamma_q$, $\gamma$, and $\beta$.
		\ELSE 
		\STATE Update $Tx_s$ and $\beta$.
		\ENDIF
		\ENDWHILE
	\end{algorithmic}
\end{algorithm}

\subsection{Missing Transactions Prediction }\label{detailpre}
To deal with the selfish and later transactions, the sending nodes predict the missing transactions at the receiving nodes and piggyback these transactions into the HCB block. The missing transaction prediction can be regarded as a binary classification problem (i.e., to classify transactions as missing or present).  

We apply naïve Bayes classifier \cite{naivebyes}, a probabilistic machine learning model, to our missing transaction predication.  Naïve Bayes classifier assumes the independence of the data's features and thus significantly simplifies the learning process. It is shown in \cite{featurenaive} that although the independence assumption is sometimes inaccurate in theory, naïve Bayes classifier has worked quite well in practice and can achieve relatively good performance for classification problems \cite{featurenaive}. The model design of the naïve Bayes classifier for our missing transaction prediction is given as follows:

\textbf{Model Design: }We denote the set containing the features of a transaction by $X=\{x_1,x_2,…,x_n\}$, where $x_i$ is the $i^{th}$ feature of the transaction. We will discuss how to select the features for transaction latter. We denote the class label of the corresponding transaction by $y_j$ and $j \in \{1,2\}$, which represents if the transaction is missing ($y_1$) or present ($y_2$) at the receiving nodes. Following the naïve Bayes classifier model, the Missing Transaction Prediction is designed to compute the a posteriori probability (APP) of each transaction's class, $p(y_j|X)$, as: 
\begin{equation}
	p({y_j}|X) = p({y_j})\mathop \prod \limits_{i = 1}^n p({x_i}|{y_j})
	\label{predictionpre}
\end{equation}
where $p(y_j)$ is the prior probability of class $y_j$; $p(x_i |y_j)$ is the conditional probability of each transaction  feature in $X$.  If the APP $p(y_j|X)$ can be computed, we can estimate the transaction's class as $y_1$ if $p(y_1|X)> p(y_2|X)$ or as $y_2$ if $p(y_1|X)< p(y_2|X)$. Therefore, when assembling an HCB block, the node can employ the features $x_i,i=1,2,\cdot \cdot \cdot,n$ of each transaction to determine whether the transaction is missing or not. To compute the probability $p(y_j|X)$,  we need to know probabilities $p(y_j)$ and $p(x_i|y_j)$, as specified in  (\ref{predictionpre}). A Naïve Bayes classifier learns to know these probabilities using a training dataset.  

\textbf{Training Phase: }With a set of transactions collected from the Ethereum MainNet (detailed in Section \ref{predictiontra}) as the training dataset, the training phase estimates the prior probability $p(y_j)$ and the conditional probabilities $p(x_i|y_j)$. Here, $p(y_j)$ is estimated by counting the frequency of training transactions that fall into class $y_j$. Similarly, $p(x_i |y_j)$ is estimated by counting the frequency of the transactions with feature $x_i$ within the training subset labeled as class $y_j$. 

\textbf{Feature Selection: }As described in Section \ref{experimentEther}, there are various complex factors that lead to the missing transactions, including network latency, transaction fee and selfish behavior from nodes. Considering these factors, we can select the following features of transactions to use in the naïve Bayes classifier model for predicting missing transaction prediction. 
\begin{itemize}
	\item Gas Fee $x_1$: It is determined by the rule of EIP-1559 \cite{EIP1559} and impacts whether the transaction is broadcast over the network. The transaction with low fee may be discarded. But in case of the transaction with high fee, it also may be deliberately withheld by some nodes. Thus, Gas fee can be selected as a feature that contributes to the prediction model. 
	
	\item Age $x_2$: It is the time duration since the sending nodes received this transaction. The transaction with the small age is likely to be not fully broadcast over the network due to the network latency. Thus, transaction age can contribute to the prediction model.
	
	\item Rank Ratio $x_3$: It is defined as the following. The transactions are sequentially packed in the block. For the $n^{th}$ transaction in a block with $m$ total transactions, its rank ratio is defined as $n/m$. The miners in Ethereum prefer to pack local transactions and the withheld transactions in the front position of their mined blocks, and thus the rank ratio relates to the selfish behavior in  our prediction model. 
	
	\item 	Existence at the sending node $x_4$: It represents whether the transaction is missing at the sending node or not (i.e., $x_4=0$ represents a missing transaction; $x_4=1$ represents a present transaction). When the transaction does not exist at the sending  node, it is likely to be missing at the receiving  nodes.
	
\end{itemize}

\subsection{HCB Forwarding Protocol}
In order to exchange blocks among HCB nodes and standard Ethereum nodes, we design a simple HCB forwarding protocol. 

\begin{figure}[tp]
	\centering
	\includegraphics[width=8.5cm]{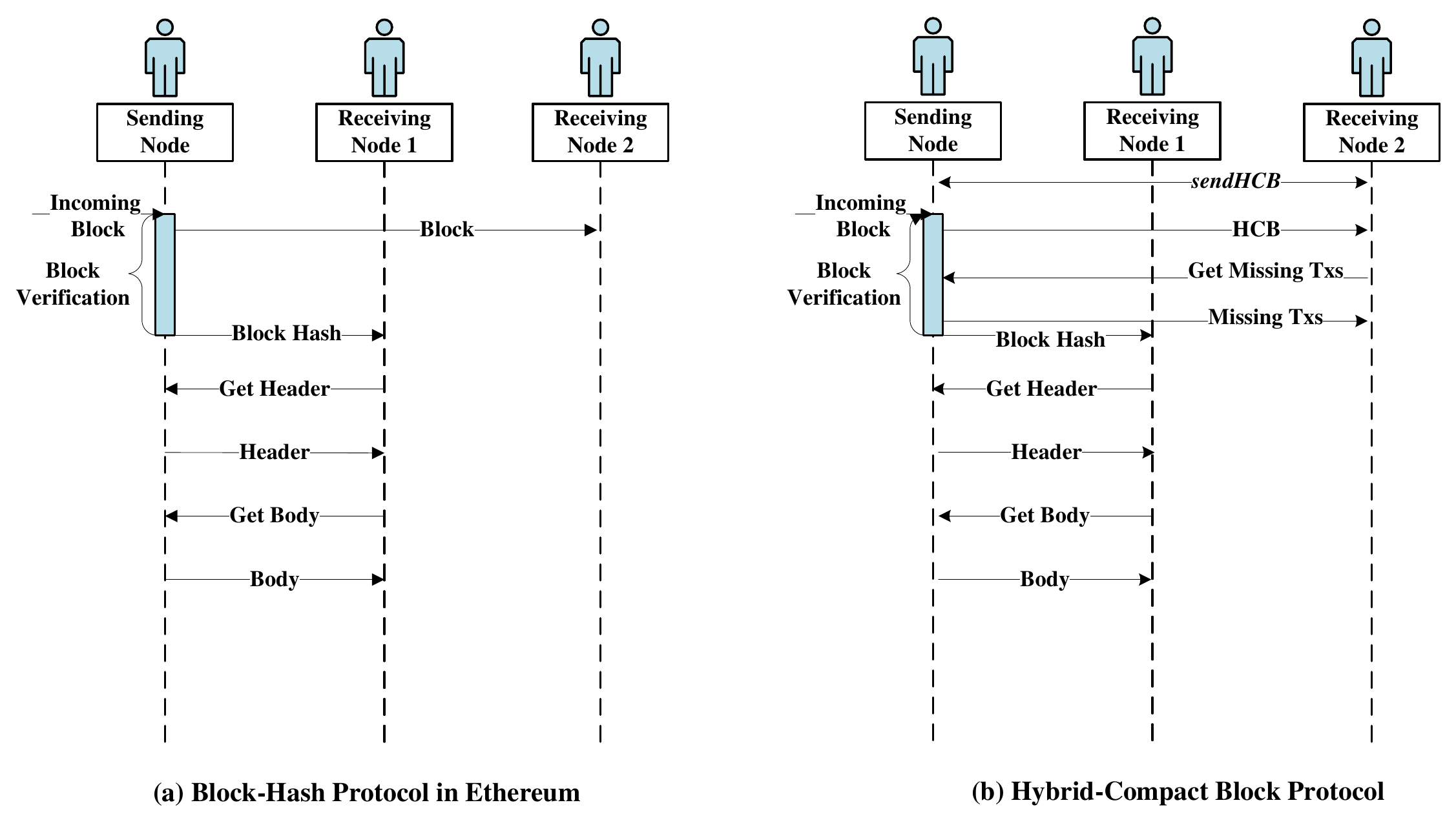}
	\caption{Block forwarding protocols: (a) Block-Hash Protocol in Ethereum (b) Our Hybrid-Compact Block Protocol.    }
	\label{bhppropa}
\end{figure}

We first briefly review the Block-Hash propagation (BHP) protocol in Ethereum, which is illustrated in Fig. \ref{bhppropa}(a). Consider that a sending node forwards a new block to its $N$ neighbor nodes. The sending node randomly selects $\sqrt N$  neighbor nodes to forward the full block directly after verifying the block head information. It then announces the block hash to the remaining neighbor nodes  after verifying the full block. The neighbor nodes  that do not have the block will request the block header and block body successively from this sending node to reconstruct the full block. 

To maintain compatibility with Ethereum, our HCB forwarding protocol is designed based on the Block-Hash propagation, as illustrated in Fig. \ref{bhppropa}(b). In particular, at the beginning when establishing a connection, the HCB nodes notify each other that they will replace the full block with an HCB block to minimize bandwidth usage through a $sendHCB$ message. When the sending node has a new block to forward, it randomly selects $\sqrt N $ neighbor nodes, and then forwards the HCB block to the selected HCB nodes and forwards the full block to the selected full nodes, accordingly. For the remaining neighbors, the block hash is forwarded, which is the same as in Fig. \ref{bhppropa}(a). For the nodes that receive an HCB block, they reconstruct the full block from Secondary Pool and Tx-Pool. If the reconstruction fails due to missing transactions, they will request the missing transactions from the sending node.

\section{Experimental Result}\label{experimentpre}
In this section, we first train and test our prediction model by crawling the data from Ethereum MainNet. We then implement an HCB prototype and evaluate its performance over the Ethereum MainNet.
\subsection{Missing Transaction Prediction}\label{predictiontra}
\textbf{Collecting Dataset:} To validate the prediction model, we set up two modified nodes with only the Secondary Pool module, and connected them to Ethereum MainNet during the period of April 8, 2022 to April 14, 2022 to collect the data. When receiving a new block from other neighbor nodes, a modified node obtained the features ($x_1$, $x_2$, $x_3$, $x_4$) of each transaction in the block and recorded them in its log file. It then generated the corresponding compact block and sent it to the other modified node. The receiving node labeled the class of transactions by querying its local Tx-Pool and the Secondary Pool, recorded the classification in its log file. After the data has been collected, we randomly select 1,000,000 transactions from the log files, and take 80\% of the selected transactions as the training dataset and 20\% of the selected transactions as the test dataset.

\textbf{Prediction Model with Training Dataset: }  By analyzing and counting the training dataset, we estimate each prior probability $p(y_j)$ and conditional probability $p(x_i|y_j)$ according to the process of the training model as in Section \ref{detailpre}. Note that both the label $y_j$ and feature $x_4$ are discrete, $p(y_j)$ and $p(x_4|y_j)$ are simply estimated by counting the corresponding frequencies. For the continuous  features (i.e., $x_1$, $x_2$ and $x_3$), their conditional probabilities are estimated by using the Curve Fitting Tool in Matlab to fit their probability density functions. 

For class $y_1$ that represents the present transactions at the receiving node and class $y_2$ that represents the missing transactions for the receiving node, the prior probabilities are given by 
\begin{equation}
	p({y_1}) = 0.92,\;\;\;p({y_2}) = 0.08
	\label{eq5}
\end{equation}

\begin{figure}[tp]
	\centering
	\includegraphics[width=8cm]{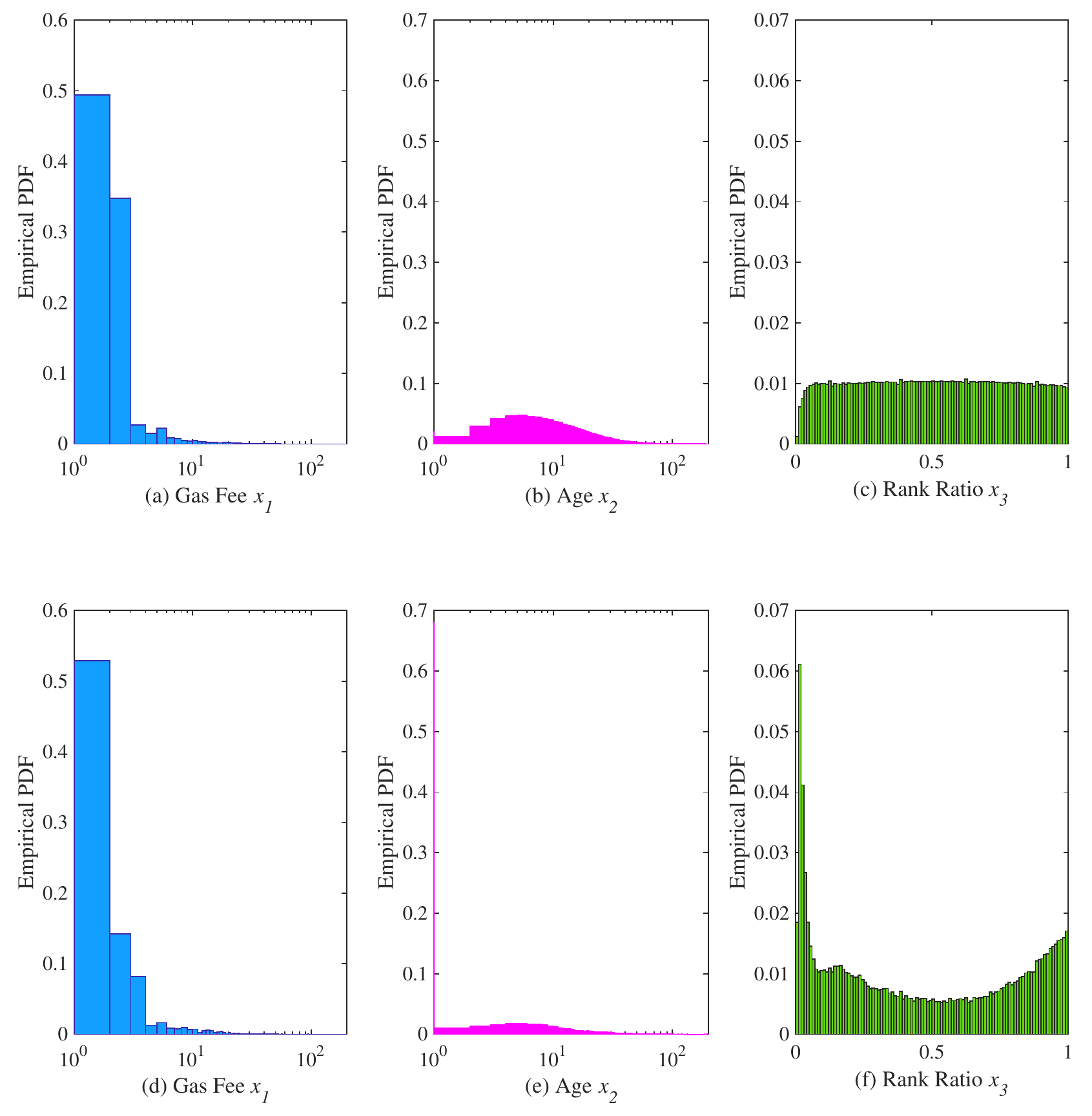}
	\caption{Frequency distributions for different continuous features in present(top) and missing (bottom) transactions subset.   }
	\label{feefit}
\end{figure}

Since fee $x_1$ is a continuous feature and the fees of most transactions take values  in the range of $[0gWei,200gWei]$, the histograms of $x_1$ under classes $y_1$, $y_2$ within the range are shown in Fig. \ref{feefit}(a) and Fig. \ref{feefit}(d), respectively. The conditional probabilities of $x_1$  given $y_1$, $y_2$ are fitted as 
\begin{equation}
	p({x_1}|{y_1}) = \frac{{0.6222}}{{x_1^2 - 0.1143{x_1} + 0.8207}}
	\label{eq6}
\end{equation}
\begin{equation}
	p({x_1}|{y_2}) = \frac{{0.5636}}{{x_1^2 - 0.2979{x_1} + 0.7611}}
	\label{eq7}
\end{equation}

Since age $x_2$ is a continuous feature and the ages of most transactions take values   in the range of $[0s,200s]$, the histograms of $x_2$  under classes $y_1$, $y_2$  within the range are shown in Fig. \ref{feefit}(b) and Fig. \ref{feefit}(e), respectively. The conditional probabilities of $x_2$  given $y_1$, $y_2$ are fitted as
\begin{equation}
	p({x_2}|{y_1}) = 0.0275{e^{ - {{(\frac{{{x_2} - 6.447}}{{5.632}})}^2}}} + 0.02145{e^{ - {{(\frac{{{x_2} - 13.44}}{{20.46}})}^2}}}
	\label{eq8}
\end{equation}
\begin{equation}
	p({x_2}|{y_2}) = \frac{{0.0306}}{{{x_2} + 0.04501}}
	\label{eq9}
\end{equation}

Since rank ratio $x_3$ is also a continuous feature, and $x_3$ for each transaction in the block is in the range of $[0,1]$, its histograms under classes $y_1$, $y_2$ are shown in Fig. \ref{feefit}(c) and Fig. \ref{feefit}(f), respectively. The conditional probabilities of $x_3$ given $y_1$, $y_2$ can be fitted as 
\begin{equation}
	p({x_3}|{y_1}) =  - 0.003665x_3^2 + 0.004119{x_3} + 0.009356
	\label{eq10}
\end{equation}
\begin{equation}
	p({x_3}|{y_2}) = \frac{{0.0047{x_3} + 0.001141}}{{{x_3} + 0.01187}}
	\label{eq11}
\end{equation}

Feature $x_4$, the existence of the transaction at the sending node or not, it has two values, missing or present. The conditional probabilities of $x_4$  given $y_1$, $y_2$ are estimated as
\begin{equation}
	p({x_4}|{y_1}) = \{ _{0.01,\;\;\;\;{x_4} = missing}^{0.99,\;\;\;\;{x_4} = present}
	\label{eq12}
\end{equation}
\begin{equation}
	p({x_4}|{y_2}) = \{ _{0.77,\;\;\;\;{x_4} = missing}^{0.23,\;\;\;\;{x_4} = present}
	\label{eq12}
\end{equation}
By substituting the above conditional probabilities and prior probabilities into eq. (\ref{predictionpre}), we can obtain the prediction model as discussed in Section \ref{predictiontra}. 

\textbf{Prediction Model Validation with Test Dataset: }We use the test dataset to validate the prediction model. In this validation, we mark the missing transactions at the receiving node as positive and the present transactions at the receiving node as negative. We then obtain the confusion matrix for the prediction results as shown in Table \ref{matrix}. By comparing the prediction results and the ground truths in Table \ref{matrix}, the precision and recall of the prediction results are calculated as $\frac{9385}{9385+483} \approx 0.951$,   $\frac{9385}{9385+12182} \approx  0.435$. In our scheme, when the sending nodes predict the transactions in the block as the missing transactions at the receiving nodes, they will send the complete transactions along with the compact block to the receiving nodes. Therefore, the precision of the prediction results determines the matched-block probability. The achieved precision of 0.951 in our experiment shows that the prediction model can efficiently predict the missing transactions at the receiving nodes. The achieved recall of 0.435 shows that more than half of the complete transactions in HCB (predicted as missing transactions) are not necessary (i.e., these transactions are already present at the receiving nodes), which means an imperfect utilization of the bandwidth. Additionally, the sending nodes would send around $\frac{9385+12182}{200000}*100\% \approx 10.7\%$ of the transactions in the corresponding compact block to the receiving nodes. 
\begin{table}[]
	\caption{Confusion Matrix for Prediction Results}
	\begin{tabular}{|c|c|c|c|}
		\hline
		\diagbox[width=8em]{Truth}{Prediction \\  results}                                                 & \begin{tabular}[c]{@{}l@{}}Positive \\ (Missing)\end{tabular} & \begin{tabular}[c]{@{}l@{}}Negative \\ (Present)\end{tabular} & Recall and Precision                            \\ \hline
		\begin{tabular}[c]{@{}l@{}}Positive \\ (Missing)\end{tabular} & 9385                                                          & 12182                                                         & \begin{tabular}[c]{@{}c@{}}Recall: \\ $\frac{9385}{9385+12182}\approx 0.435$\end{tabular}  \\ \hline
		\begin{tabular}[c]{@{}l@{}}Negative \\ (Present)\end{tabular} & 483                                                           & 177950                                                        & \begin{tabular}[c]{@{}c@{}}Precision: \\ $\frac{9385}{9385+483}\approx 0.951$\end{tabular} \\ \hline
	\end{tabular}
	\label{matrix}
\end{table}
\subsection{Performance of HCB on Ethereum MainNet}
In this subsection, we measure the performance of HCB on Ethereum MainNet, in terms of matched-block probability, matched-transaction rate, block size distribution, and block propagation time. 

\textbf{Experiment Setup: }To ensure the performance of HCB on Ethereum MainNet, we implemented the prototype of the full HCB protocol with the Secondary Pool and the prediction module based on current Ethereum software \cite{GoEthereum}. For comparison, we also implemented three compact block-like protocols: 1) the basic compact block (BCB); 2) the simple compact block only with Secondary Pool (SCB); 3) the compact block only with the prediction module (PCB). We conducted four sets of experiments to measure the performances of the five protocols (i.e., BHP, HCB, BCB, SCB, PCB) during the period of August 7, 2022 to September 10, 2022. In each set of experiments, we setup two AliCloud servers; in different set of experiment, two servers were at different geographical locations; in all sets of experiments, all servers were configured with the same hardware configuration as described in Section \ref{experimentsta}. The locations of the servers in each set are given as follows: Shenzhen and Shanghai in the first set, Shenzhen and Sydney in the second set, Shenzhen and London in the third set, and California and Mumbai in the last set. In each set of experiments, two nodes that run with the same Ethereum Client software were deployed onto the two servers respectively and connect on Ethereum MainNet; the two nodes adopted one of the five protocols for one week; during the experiment, the two nodes exchanged the blocks and transactions from other standard nodes on Ethereum MainNet, and forwarded the corresponding messages to each other according to the adopted protocol.

\textbf{Matched-Block Probability and Matched-Transaction Rate: }We calculate the matched-block probability and transaction-matched rate of the four compact block protocols (BCB, SCB, PCB, and HCB). Matched-transaction rate is defined as the ratio of the present transaction number to the total transaction number in one compact block, and it is an important metric to demonstrate the performance of compact-block-like protocols. 

\begin{figure}[tp]
	\centering
	\includegraphics[width=7cm]{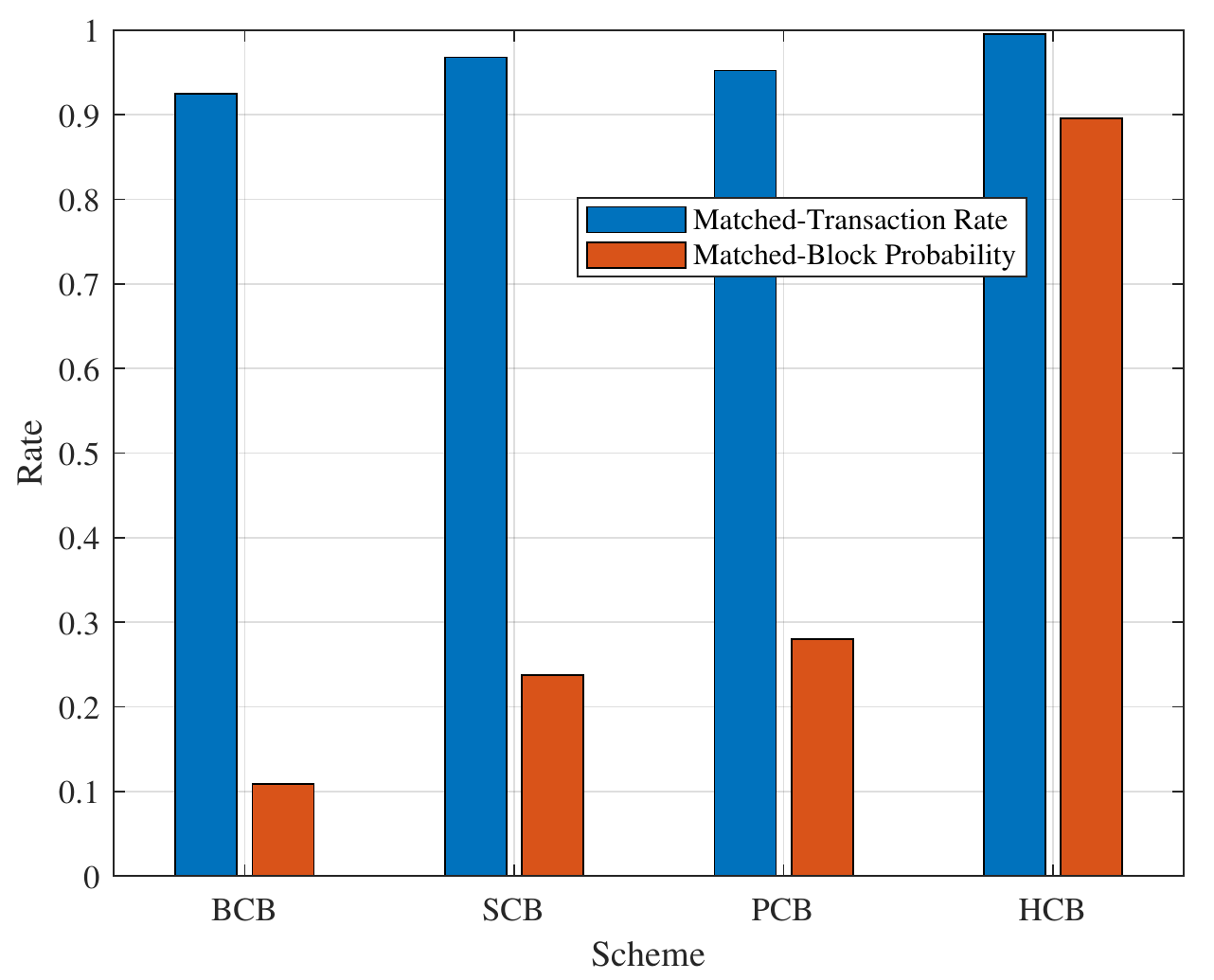}
	\caption{Matched-Block probability (red columns) and Matched-Transaction rate (blue columns) under different schemes.    }
	\label{matchedblock}
\end{figure}
The experiment results are shown in Fig. \ref{matchedblock}. We can see that the matched-block probability of BCB is only around 0.12 if we directly adopt the basic compact block in Ethereum, although the matched-transaction rate can reach around 0.92. This means that the nodes can match most of the transactions in a block, but at the same time there is a large chance that the nodes need extra communication rounds to request the missing transactions. And if we merely add the Secondary Pool (SCB protocol) or the prediction model (PCB protocol) to BCB, the matched-block probability can be improved to 0.23 and 0.28 respectively. This is because merely adding the Secondary Pool cannot match the selfish transactions that are not broadcast over the network, and merely adding the prediction model cannot fully predict all missing transactions (Table \ref{matrix} shows that the precision of prediction results is around 0.951). For our HCB protocol, the matched-block probability can be significantly improved to around 0.90 (for comparison, the matched-block probability in Bitcoin is around 0.85 \cite{churn}), and the matched-transaction rate can be improved to around 0.99. Therefore, HCB nodes are likely to reconstruct the full blocks locally without requesting the missing transactions using extra communication rounds and thus can speed up the block propagation significantly in Ethereum.

\begin{figure}[tp]
	\centering
	\includegraphics[width=7cm]{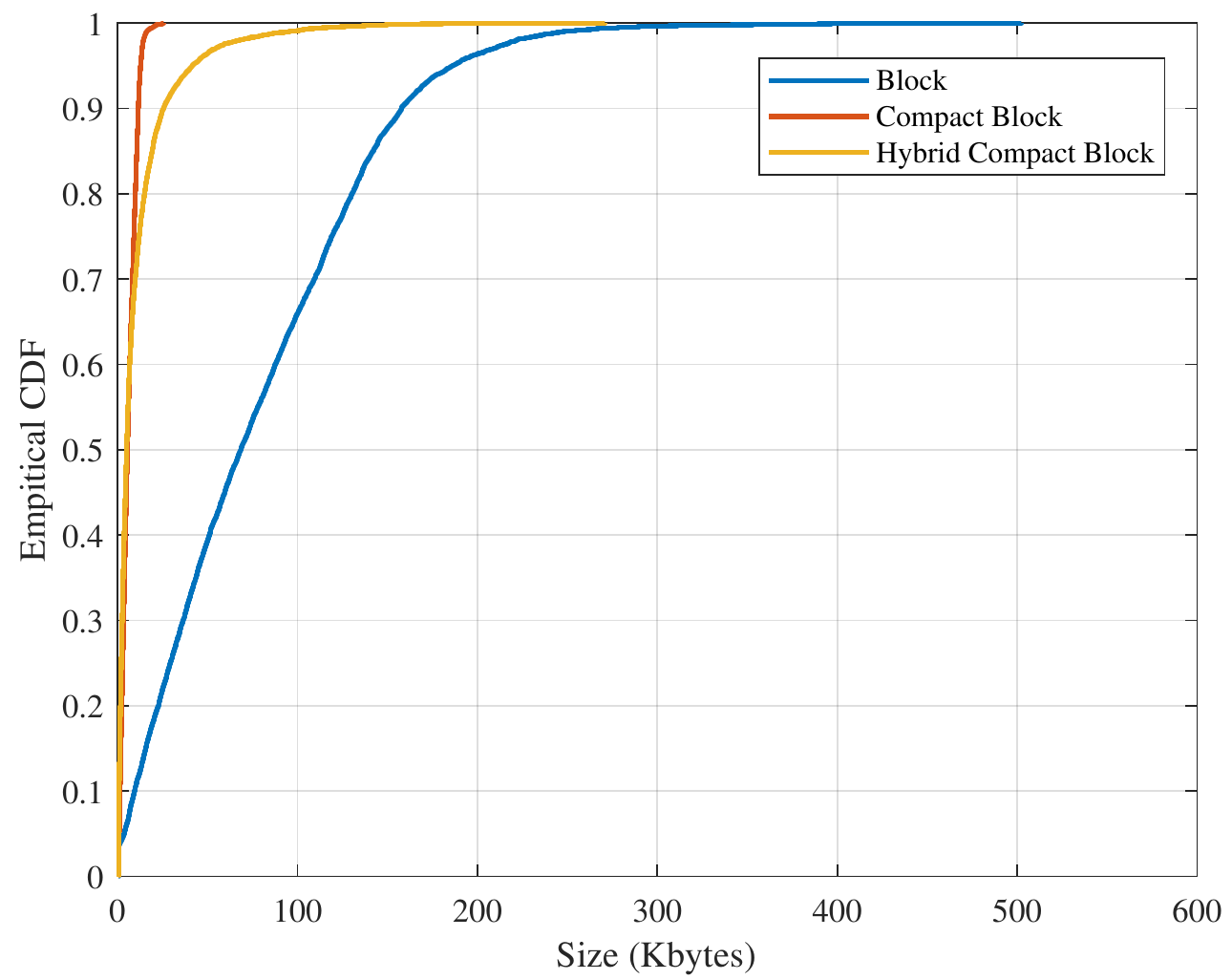}
	\caption{The CDFs of the size of block (blue), compact block (red) and hybrid-compact block (yellow) in Ethereum.   }
	\label{hcbsize}
\end{figure}

\textbf{Block Size: }Our HCB and other compact block protocols speed up the block propagation by reducing the block size. We recorded the size of full block, compact block and hybrid-compact block during the experiment, and their CDFs are shown in Fig. \ref{hcbsize}. We can see that around 80\% of the full blocks are larger than 20 Kbytes. Works \cite{information,scale} have demonstrated that the block propagation time increases quickly when the block size exceeds 20 Kbytes. In BCB  , only 0.5\% of compact blocks are larger than 20 Kbytes, but the probability of retransmitting missing transactions is as high as 90\% (i.e., the matched-block probability of BCB is around 10\% as shown in Fig. \ref{matchedblock}); in HCB, 15\% of hybrid-compact blocks are larger than 20 Kbytes since some full transactions are inserted into the block, and the retransmission probability is decreased to 10\%. 

\textbf{Block Propagation Time: }Block propagation time is the most important metric for blockchains, since it determines the fork rate of a blockchain \cite{fork1,bloxroute}. For BHP, the block propagation time is dominated by the transmission time of the full blocks. For BCB, SCB, PCB, and HCB, the block propagation time ca n be written as
\begin{equation}
	t = \rho {t_x} + (1 - \rho )({t_x} + {t_y}) = {t_x} + (1 - \rho ){t_y}
	\label{comtime}
\end{equation}
where $t_x$ is the transmission time of the initial compact block; $t_y$ is the retransmission time of the missing transactions; $\rho $ is the expected matched-block probability of the protocol; $(1 - \rho ){t_y}$ is the additional retransmission time of missing transactions of the protocol. 

\begin{figure}[tp]
	\centering
	\includegraphics[width=8.5cm]{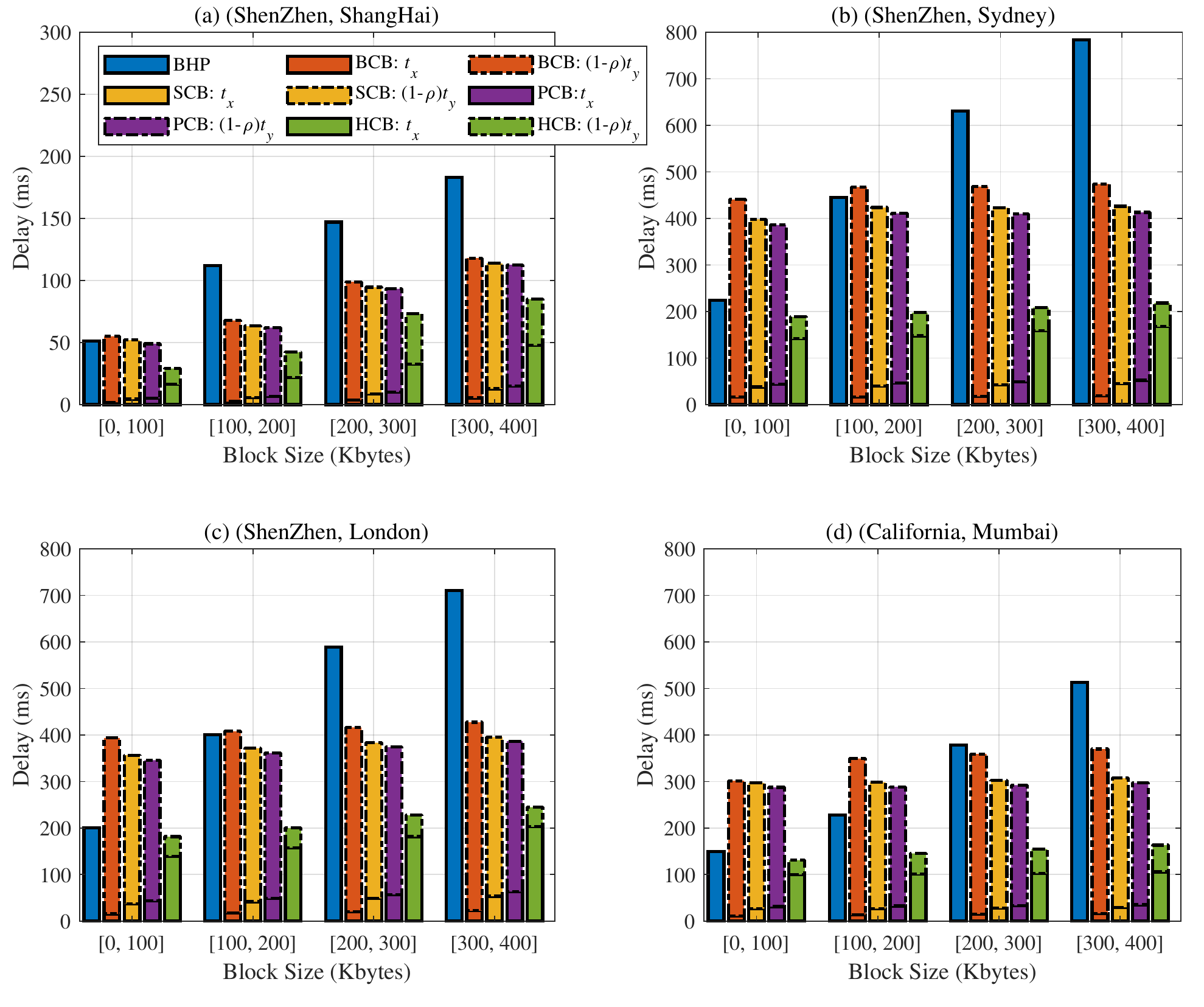}
	\caption{Block propagation time for different schemes between two nodes at different locations (a) Shenzhen and Shanghai; (b) Shenzhen and Sydney; (c) Shenzhen and London; (d) California and Mumbai.   }
	\label{hcbpropa}
\end{figure}

The experiment results of block propagation time are shown in Fig. \ref{hcbpropa}, where the horizontal axis indicates the block size. From the results in Fig. \ref{hcbpropa}, we can make several conclusions as follows. For our HCB, it saves more than 1/2 block propagation time compared to the BHP protocol when the block size is large. Thanks to its good matched-block probability and smaller hybrid-compact block size, HCB always needs the shortest block propagation time among all protocols under any block size and physical distance between the nodes. BHP does not work well when the block size is large, and its block propagation time increases linearly with the block size, and it performs worst among the five protocols when the block size is larger than 200Kbytes. BCB, SCB, and PCB do not work well when the block size is small, and their block propagation times are mainly dominated by the retransmission time of the missing transactions due to their low matched probabilities. As shown in Fig. \ref{hcbsize}, currently 65\% and 95\% of blocks in Ethereum are smaller than 100Kbytes and 200KBytes respectively, which hinders the adoption of BCB, SCB, and PCB in Ethereum-like blockchains. 

\section{Conclusion and Discussion}\label{conclusion}
Compact block is a good way to speed up block propagation in blockchain such as Bitcoin. But due to the smaller transaction pool, network latency, and the selfish behaviors of nodes, the basic compact block cannot be directly adopted in the Ethereum blockchain. This work proposes a hybrid-compact block (HCB) propagation protocol with two new modules, the Secondary Pool and Missing Transaction Prediction, which improve the matched-block probability of HCB in Ethereum from 12\% to 90\% so that the block propagation time can be shortened significantly. As a result, the propagation time of HCB only needs 1/3$\sim$1/2 time of the current protocol in Ethereum when the block size is large. Moreover, the implementation of the two modules is only an internal extension to the standard Ethereum node architecture, and HCB is compatible with the current Ethereum network. 

Our HCB is a block propagation scheme, and it is compatible with different Ethereum-like blockchains, such as Ethereum 2.0 \cite{ETHPOS}, Ethereum Classic (ETC) \cite{ETHClass}, Ethereum-PoW (ETHW) \cite{ETHW}. Ethereum 2.0 replaces the current Ethereum's Proof-of-Work (PoW) with Proof-of-Stake (PoS) \cite{casper,ghostcasper} at the consensus layer. But the node in Ethereum 2.0 also maintains the Tx-Pool module to store the unconfirmed transactions, and the  proposer (miner in PoW) selects the eligible transactions from the Tx-Pool to assemble the block and broadcast it over the network. Both the Secondary Pool and Missing Transaction Prediction modules can be added in the architecture of Ethereum 2.0 to support HCB. Moreover, it is more beneficial to apply HCB in Ethereum 2.0 to speed up the block propagation and reduce the network load. Unlike current Ethereum (that adopts PoW), Ethereum 2.0 has a fixed block interval of 12 seconds, short block propagation time allows more time for validation, cryptographic operations, state, and ledgers management \cite{ETHPOS,gossipp2p}. 

Although the presented HCB performs well, its detailed algorithms can be further improved. For example, the Naïve Bayes Classifier assumes independence features, which is not accurate in our missing transaction prediction model and leads to a probability of 10\% to retransmit the missing transactions. More sophisticated classifiers, such as the Semi-Naïve Bayes Classifier \cite{seminaive}, Bayesian Network \cite{bayesiannetwork}, Random Forest \cite{randomForest}, Logistic Regression \cite{logisticRegression}, and Support Vector Machine \cite{supportvector}, are good options for improving our prediction model. 
\ifCLASSOPTIONcaptionsoff
\newpage
\fi

\bibliographystyle{IEEEtran}

\bibliography{refs}

\end{document}